\documentclass[aps,pr,twocolumn,superscriptaddress,floatfix,longbibliography]{revtex4-1}

\usepackage{physics, amsmath,amssymb, siunitx}
\usepackage{graphicx}
\usepackage{booktabs}
\usepackage{color}
\usepackage{xcolor}
\usepackage{enumitem}
\usepackage[normalem]{ulem}
\setlist{noitemsep,leftmargin=*}
\usepackage{hyperref}
\hypersetup{
    colorlinks=true,
    urlcolor= blue,
    citecolor=blue,
    linkcolor= blue,
    bookmarksopen=false,
    }
\usepackage[titletoc,toc,title]{appendix}

\newcommand{\black}[1]{{\textcolor{black}{#1}}}

\newcounter{para}

\def \smb{SmB$_6$ }
\def \smbp{SmB$_6$}

\def \oto{$1 \times 1$ }
\def \tto{$2 \times 1$ }

\begin{document}

\title{Consistency between ARPES and STM measurements on \smb}

\author{Christian E. Matt}
\affiliation{Department of Physics, Harvard University, Cambridge, MA, 02138, USA}
\author{Harris Pirie}
\affiliation{Department of Physics, Harvard University, Cambridge, MA, 02138, USA}
\author{Anjan Soumyanarayanan}
\affiliation{Department of Physics, Harvard University, Cambridge, MA, 02138, USA}
\author{Yang He}
\affiliation{Department of Physics, Harvard University, Cambridge, MA, 02138, USA}
\author{Michael M. Yee}
\affiliation{Department of Physics, Harvard University, Cambridge, MA, 02138, USA}
\author{Pengcheng Chen}
\affiliation{Department of Physics, Harvard University, Cambridge, MA, 02138, USA}
\author{Yu Liu}
\affiliation{Department of Physics, Harvard University, Cambridge, MA, 02138, USA}
\author{Daniel T. Larson}
\affiliation{Department of Physics, Harvard University, Cambridge, MA, 02138, USA}
\author{Wendel S. Paz}
\affiliation{Departamento de F\'{i}sica, Universidade Federal do Esp\'{i}rito Santo (UFES), Av. Fernando Ferrari, 514, 29075-910, Vit\'{o}ria, ES, Brasil.}
\affiliation{Instituto de F\'{i}sica, Universidade Federal do Rio de Janeiro, Caixa Postal 68528, Rio de Janeiro, RJ 21941-972, Brazil}
\author{J.J. Palacios}
\affiliation{Departamento de F\'{i}sica de la Materia Condensada, Condensed Matter
Physics Center (IFIMAC) and Instituto Nicol\'{a}s Cabrera (INC), Universidad Aut\'{o}noma de Madrid, Cantoblanco, 28049 Madrid, Spain}
\author{M.H. Hamidian}
\affiliation{Department of Physics, Harvard University, Cambridge, MA, 02138, USA}
\author{Jennifer E. Hoffman}
\affiliation{Department of Physics, Harvard University, Cambridge, MA, 02138, USA}
\date{\today}

 \begin{abstract}


The Kondo insulator SmB$_6$ has emerged as a primary candidate for exotic quantum phases, due to the predicted formation of strongly-correlated, low-velocity topological surface states, and corresponding high Fermi-level density of states. 
However, measurements of the surface-state velocity in SmB$_6$ differ by orders of magnitude, depending on the experimental technique used. Here we reconcile two techniques, scanning tunneling microscopy (STM) and angle-resolved photoemission spectroscopy (ARPES), by accounting for surface band bending on polar terminations. Using spatially-resolved scanning tunneling spectroscopy (STS), we measure a band shift of $ \sim 20$ meV between full-Sm and half-Sm terminations, in qualitative agreement with our density functional theory (DFT) calculations of the surface charge density. Furthermore, we reproduce the apparent high-velocity surface states reported by ARPES, by simulating their observed spectral function as an equal-weight average over the two band-shifted domains that we image by STM.
Our results highlight the necessity of local measurements to address inhomogeneously-terminated surfaces, or fabrication techniques to achieve uniform termination for meaningful large-area surface measurements of polar crystals such as SmB$_6$.
\end{abstract}
 
\maketitle


\section{Introduction}

\begin{figure}[hb]
\centering
\includegraphics[width=\columnwidth]{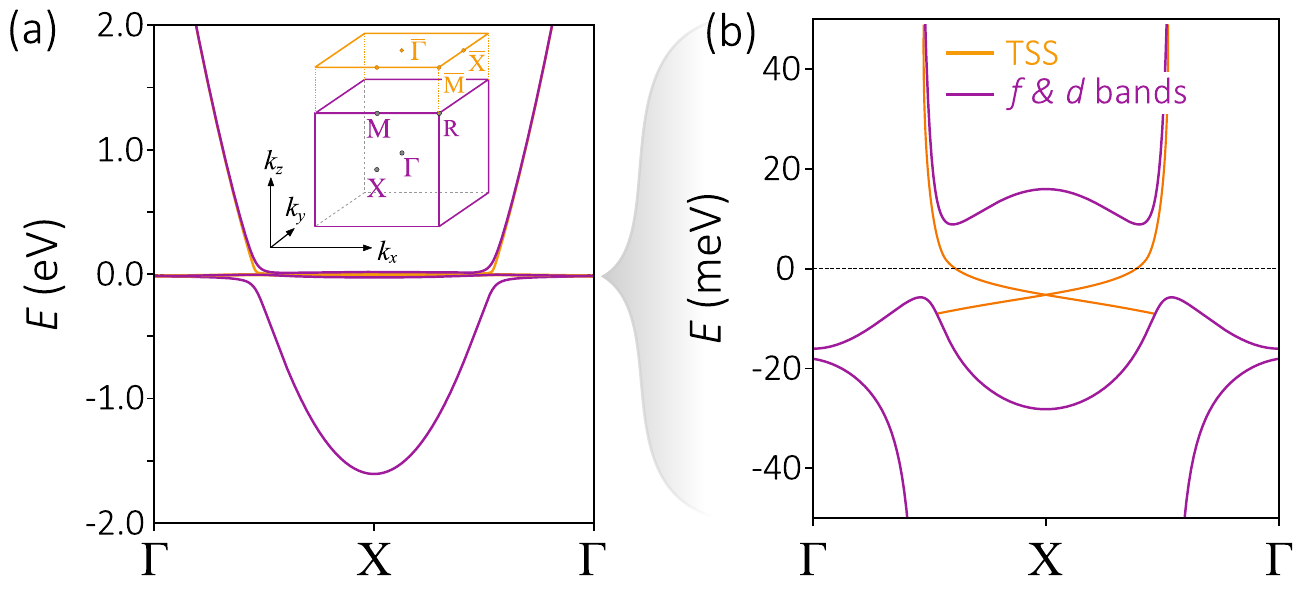}
\caption{(a) Schematic of the \smb band structure, showing two heavy $f$ bands hybridizing with a light $d$ band (all purple), and topological surface states (TSS, orange) that have a low velocity. Inset of (a): Bulk and surface Brillouin zone of \smbp. (b) Zoom-in of band structure at the Fermi level.}
\label{fig:cartoon}
\end{figure}

    
In a Kondo insulator (KI), strong interactions between localized $f$ electrons renormalize their spectral weight towards the chemical potential. Below a characteristic temperature $T^*$, conduction electrons begin to scatter from these renormalized $f$ states, opening a hybridization gap at the Fermi level. In a subset of KIs called topological Kondo insulators, this gap can encode a non-trivial bulk topological invariant, leading to the appearance of protected surface states \cite{MDzeroARCMP2016,PMisra2008}. In the KI \smbp, the onset of the hybridization gap leads to a resistivity upturn below $\sim50$ K \cite{DJKimNatMat2014, JAllenPRB1979,JCooleyPRL1995}. 
Yet, rather than diverging, the resistivity saturates below 5 K, indicating the emergence of an additional conduction channel \cite{WolgastPRB2013, SyersPRL2015}. This conduction channel has been attributed to topological surface states by several theoretical studies, which span complementary approaches including renormalized band theory and tight-binding Hamiltonians matched to LDA (+Gutzwiller) calculations \cite{MDzeroPRL2010, FLuPRL2013, alexandrov2013cubic}. These calculations predict the existence of three surface Dirac cones with heavy quasiparticles, of predominantly $f$ character, as shown schematically in Fig.\ \ref{fig:cartoon}. 
Such low-velocity Dirac fermions would provide a high density of states at the Fermi level, increasing their susceptibility to exotic orders and their potential utility~\cite{alexandrov2013cubic, chenPhysRevB2014a,efimkinPhysRevB2014,thomsonPhysRevB2016a}.
However, the empirical identification of the additional conduction channel \cite{WolgastPRB2013, SyersPRL2015} with the predicted topological surface states \cite{MDzeroPRL2010, FLuPRL2013, alexandrov2013cubic} has remained controversial due to apparent contradictions between different experimental techniques.

%
%

Experimentally, angle-resolved photoemission spectroscopy (ARPES) and scanning tunneling microscopy (STM) have each identified key features of the topological states in \smbp, but with quantitative and qualitative differences.  At low temperatures, ARPES studies reported a hybridization gap that hosts linearly dispersing surface states \cite{JDenlingerJPSJC2014, NXuPRB2013, NXuPRB2014, MNeupaneNATCOMM2013, JJiangNATCOMM2013} with a non-trivial spin texture \cite{NXuNATCOMM2014, SSugaJPSJ2014}. 
\begin{table*}[t]
\caption{\label{tab:surf} Comparison of \smb surface-state properties predicted by theory and measured by STM, ARPES, and quantum oscillations. We tabulate values for the Fermi velocity $v$, Dirac-point energy $E_D$, and surface Fermi wavevector $k_F$, at both the $\overline{\mathrm{X}}$ and $\overline{\Gamma}$ points of the surface Brillouin zone.}
  \begin{ruledtabular}
\begin{tabular}[c]{lcccc}
  & Theory \cite{FLuPRL2013}  & STM \cite{pirieNatPhys2020} & ARPES \cite{JJiangNATCOMM2013} & \black{Quantum Oscillation} \cite{liScience2014}
\\ \hline
$\hbar v_{\bar{X}}$ (meV$\cdot$\AA)  &  $7.6 \pm 0.3$ & $16 \pm 2$  &  $240 \pm 20$ & \black{$1900 \pm 300$}
\\
$E_{D_{\bar{X}}}$ (meV)  & $-5.4 \pm 0.1$  & $1 \pm 1$   &  $-65\pm 4$  &  \black{$-57 \pm 9$}  
\\
$(k_{F_{\bar{X}}}-\bar{X}) (\pi / a_0)$ & $0.44 \pm 0.06$ &  $0.19 \pm 0.02$    &  $0.51 \pm 0.03$ ($\Gamma - \mathrm{X} - \Gamma$) &  \black{$0.039 \pm 0.003$} 
\\
\hline
$\hbar v_{\overline{\Gamma}}$ (meV$\cdot$\AA) &  $90 \pm 9$  & $50 \pm 2$  &  $220 \pm 20$ & \black{$4300 \pm 100$ } 
\\
$E_{D_{\overline{\Gamma}}}$ (meV)  & $-9 \pm 2$  & $-7\pm 1$  &  $-23\pm 3$  &  \black{$-460 \pm 20$  }
\\
$k_{F_{\overline{\Gamma}}} (\pi/ a_0)$   & $0.07 \pm 0.01$ & $0.14 \pm 0.02$    & $0.15 \pm 0.03$ &   \black{$0.142 \pm 0.001$ }
\\
\end{tabular}
\end{ruledtabular}
\end{table*}
However, the apparent velocity of these states is an order of magnitude higher than theoretically predicted (see Table~\ref{tab:surf}). Meanwhile, the hallmark of a topological surface state---its Dirac point---has not been clearly resolved in any ARPES experiment to date \cite{NXuPRB2014}, leading to the suggestion that it has been pushed into the valence band by a strong surface potential \cite{roy2014surface}, or by the breakdown of the Kondo effect at the surface \cite{VAlexandrovPRL2015}.  On the other hand, milliKelvin scanning tunneling spectroscopy (STS) studies identified several strong resonances within the hybridization gap, consistent with low-velocity surface states \cite{JiaoNatComm2016, ZSunPRB2018}. Additionally, momentum-resolved STM directly imaged linearly dispersing low-velocity surface states that converge to a Dirac point within the gap~\cite{pirieNatPhys2020}, consistent with theoretical predictions~\cite{FLuPRL2013}.

The apparent inconsistencies between STM and ARPES arise from the different experimental length scales for each technique. STM typically images hundred-nanometer regions with picometer spatial resolution. On \smbp, STM universally observes surface domains with sizes on the order of tens of nanometers \cite{WRuanPRL2014, SRosslerPNAS2014, JiaoNatComm2016, ZSunPRB2018,pirieNatPhys2020,MYeeARXIV2013}, consistent with its polar structure and the lack of a natural cleavage plane. Yet the typical ARPES spot size is on the order of tens of microns \cite{NXuJPCM2016}, and consequently averages over thousands of \smb surface domains.  This averaging poses a problem if the various domains exhibit polarity-driven band bending, as ARPES spectra will contain a superposition of spectral features, shifted in energy with respect to one another.

Here we use STM spectroscopy to guide a simulation of the spectral functions on polar Sm \oto and non-polar Sm \tto terminations, using the energy and momentum broadening of typical ARPES experiments.
For a range of realistic experimental parameters, our \textit{simulated} ARPES spectra show topological surface states with an artificially enhanced Fermi velocity and a buried Dirac point, similar to published \textit{experimental} ARPES results. Our findings provide the long-sought, fully-consistent explanation for the apparent discrepancy between the band structure measured by ARPES and STM. They further confirm the consistency between STM and theoretical predictions of low-velocity surface states with an in-gap Dirac point and high density of states at the Fermi level.

\section{Methods}
\subsection{Scanning tunneling microscopy/spectroscopy}
 We performed STM experiments on single crystals of \smb grown using the Al-flux method \cite{nakajimaNatPhys2016, DJKimSciRep2013}. We cleaved the crystals in cryogenic ultra-high vacuum at $\sim30$ K before inserting them into the STM head. We prepared PtIr STM tips by \textit{ex situ} mechanical sharpening then \textit{in situ} field emission on Au foil.
 
\subsection{Calculations}
We performed calculations in the framework of density functional theory (DFT), as implemented in the Quantum ESPRESSO package \cite{GPaoloJPCM2009}. We calculated the exchange-correlation functional using the generalized gradient approximation of Perdew-Burke-Ernzerhof (GGA-PBE) \cite{JPPerdewPRL1996}. 
The electron-ion interactions are described by ultrasoft pseudopotentials with valence electron configurations of $2s^{2}2p^{1}$ for B atoms and $5s^{2}4d^{10}5p^{6}6s^{2}4f^{6}$ for Sm atoms. The energy cutoff for the plane wave basis is 120 Ry with a charge density cutoff of 500 Ry. We used a Monkhorst-Pack \cite{monkhorst1976special} scheme with a $12\times12\times1$ \textit{k}-mesh for the Brillouin zone integration for the supercell with one unit cell (\oto Sm) and $6\times12\times1$ $k$-mesh for the supercell with two unit cells (\tto Sm). In all calculations, the lattice parameter was fixed at the experimental value $a_0 = 4.13$ \AA, with slab thickness $20.65$ \AA\ and vacuum thickness 15 \AA\ to minimize interactions between the periodic images. We did not consider spin polarization or spin-orbit coupling since our focus is on the electrostatics of the material.

\begin{figure}[t!]
\centering
\includegraphics[width=0.98\columnwidth]{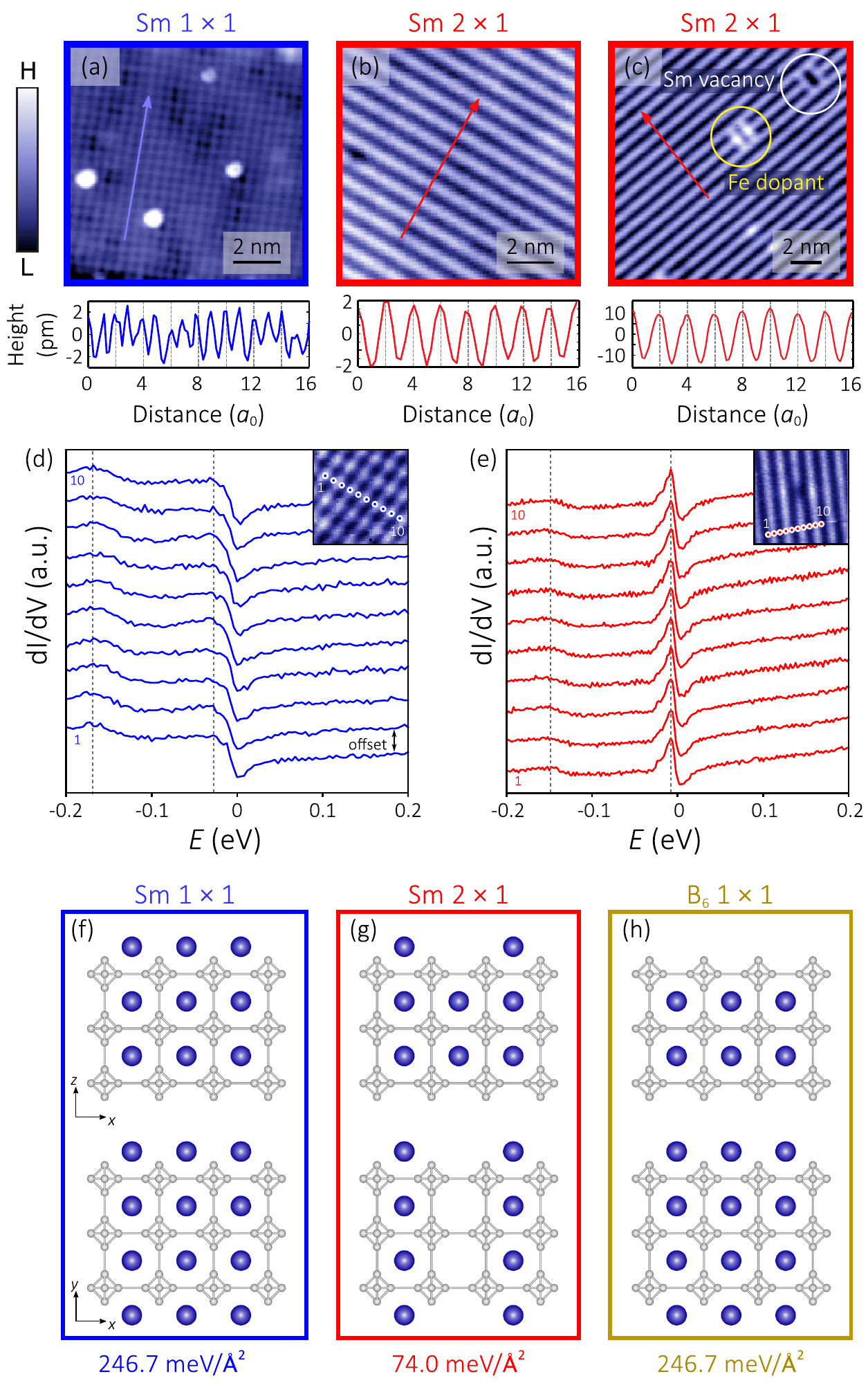}
\caption{STM topography of the (a) Sm \oto termination and the Sm \tto termination of (b) pristine \smb and (c) Fe-doped \smb~\cite{pirieNatPhys2020}.  Acquisition parameters are: (a) $T = 9.5$ K, $V_s = -200$ mV, $R_J = 10\; \mathrm{G}\Omega$ (b) $T = 8.5$ K, $V_s = -100$ mV, $R_J = 5\; \mathrm{G}\Omega$ and (c) $T = 6.5$ K, $V_s = -100$ mV, $R_J = 0.5\; \mathrm{G}\Omega$. (d) - (e) Spatially homogeneous $dI/dV$ spectra on the Sm \oto and Sm \tto surface. Each curve is offset for clarity. The location is indicated in the inset of each panel. The inset in (d) shows an area of $2.5 \times 2.5$ nm$^2$ and (e) an area of $5.1 \times 5.1$ nm$^2$. 
Acquisition parameters are: (d) $T=9.5$ K, $V_\mathrm{s}=-200$ mV, $R_\mathrm{J} = 2$ G$\Omega$, bias excitation amplitude $V_\mathrm{rms} = 2.82$ mV, and (e) $T=6.5$ K, $V_\mathrm{s}=200$ mV, $R_\mathrm{J} = 1$ G$\Omega$, bias excitation amplitude, $V_\mathrm{rms}=1.41$ mV. (f) - (h) Side-view (upper) and top-view (lower) of different surface terminations and their corresponding formation energies, calculated by DFT.}
\label{fig:surface}
\end{figure}
\section{Results}
\subsection{Surface characterization}
Due to its lack of a natural cleavage plane, an abundance of distinct surface terminations have been observed by STM on \smb \cite{SRosslerPM2016}. Across a dozen STM experiments, the largest reported domain of an ordered surface on pristine \smb ($<1\%$ dopants) is only 60 nm \cite{MYeeARXIV2013, WRuanPRL2014, SRosslerPNAS2014, JiaoNatComm2016, ZSunPRB2018,pirieNatPhys2020, SRosslerPM2016,jiaoSciAdv2018, herrmannArXiv2018}. 
Two commonly observed surfaces are the \oto square lattice [Fig.\ \ref{fig:surface}(a)] and the \tto rows that arise when half of the Sm atoms are removed during cleaving [Fig.\ \ref{fig:surface}(b) and \ref{fig:surface}(c)] \cite{SRosslerPM2016, repository}.
The \tto surface has also been observed by 
low-energy electron diffraction \cite{ramankuttyJESP2016} and ARPES, where it manifests as Umklapp scattering \cite{PHlawenkaNATCOMM2018,NXuPRB2013}.
We confirmed the identity of the \tto surface using lightly Fe-doped samples where Fe is known to substitute for Sm \cite{KAkintolaPRB2017}; we observed individual Fe-atom signatures centered on the rows of Sm atoms in Fig.\ \ref{fig:surface}(c). We confirmed the identity of the \oto lattice presented in Fig.\ \ref{fig:surface}(a) as a full Sm layer due to the direction of its band bending compared to the \tto surface, as shown in Fig.\ \ref{fig:surface}(d-e) and discussed in more detail below. 

The relative prevalence of each surface can be understood from its formation energy [Figs.\ \ref{fig:surface}(f-h)]. Although most STM reports have focused on the \oto surface \cite{JiaoNatComm2016, SRosslerPNAS2014, ZSunPRB2018, WRuanPRL2014}, our more frequent observation of the \tto surface is consistent with its lower formation energy as calculated by DFT. In general, a more balanced charge distribution on either side of the cleave, as drawn in Fig.\ \ref{fig:surface}(g), is intuitively expected to lower the surface formation energy.

\begin{figure}[t]
\centering
\includegraphics[width=\columnwidth]{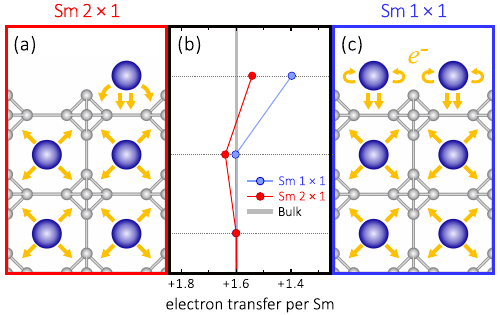}
\caption{DFT-calculated electron transfer from Sm atoms to B$_6$ clusters for the \tto surface (a) and the \oto surface (c). Fewer electrons are drawn from each Sm atom on the \oto surface as compared to the \tto surface. } 
\label{fig:charge-transfer}
\end{figure}

\begin{figure*}[t]
\centering
\includegraphics[width=\textwidth]{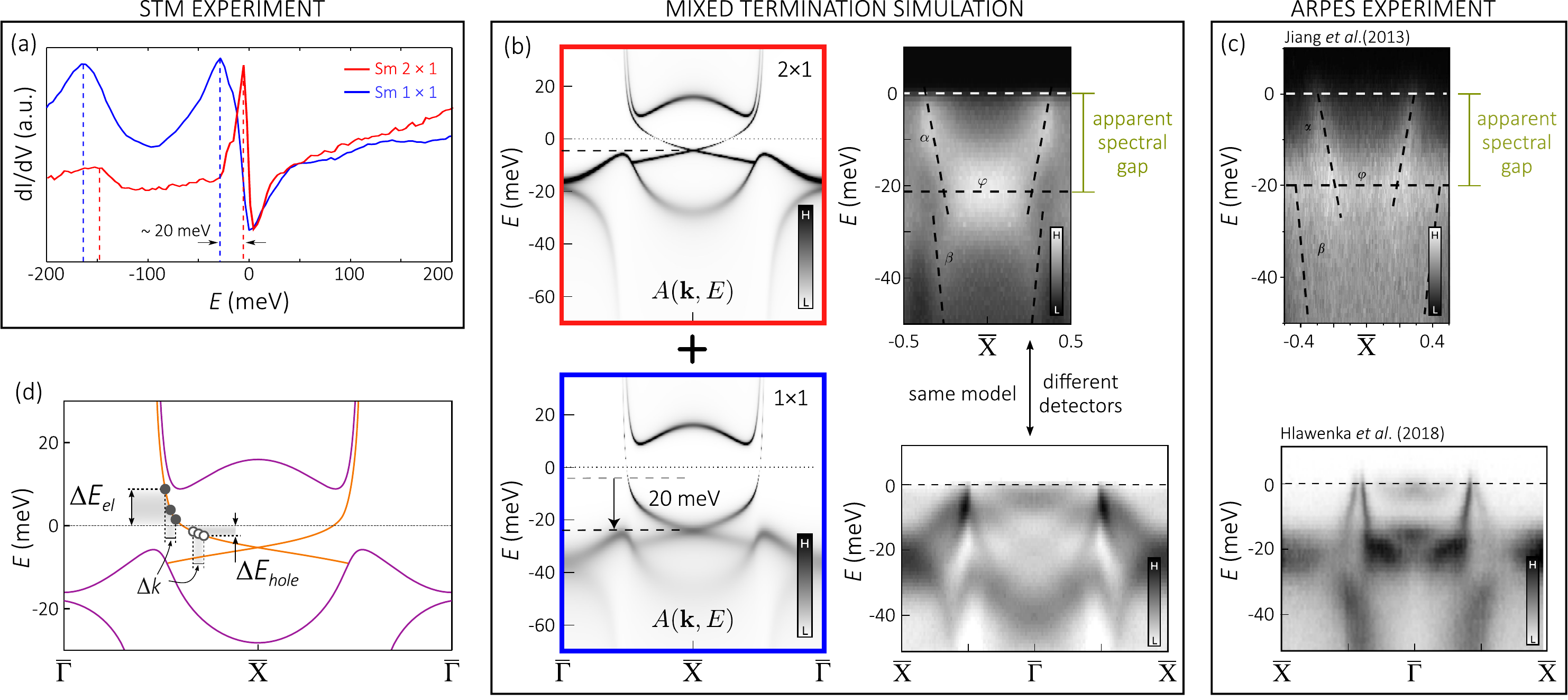}
\caption{(a) Measured $dI/dV$ on two different surfaces of \smbp. Acquisition parameters are (blue curve) $T = 9$ K, $V_s = -250$ mV, $R_J = 2$ G$\Omega$, bias excitation amplitude $V_\mathrm{rms}=2.8$ mV, and (red curve) $T = 8$ K, $V_s = 200$ mV, $R_J = 1$ G$\Omega$, $V_\mathrm{rms} = 1.4$ mV. (b) Starting with the electronic structure derived by STM on the non-polar \tto surface (red) \cite{pirieNatPhys2020}, we inferred the electronic structure on the \oto polar termination by rigidly shifting the occupied states down by $20$ meV (blue), based on our local STS measurements. The average of the simulated spectral functions from the \tto and \oto surfaces imitates the result of a spatially averaging measurement such as ARPES. We convoluted the averaged spectral function with a Gaussian kernel in order to account for finite temperature, energy and momentum resolution. The following realistic experimental parameters
 have been used to simulate the spectra along the $\mathrm{\overline{M}-\overline{X}-\overline{M}}$ and $\mathrm{\overline{X}-\overline{\Gamma}-\overline{X}}$ directions. Upper panel: $T=12$ K, $\Delta E = 10$ meV, $\Delta k=0.04$ \AA$^{-1}$ (as reported in Ref.\ \cite{JJiangNATCOMM2013}); lower panel: $T=1$ K, $\Delta E = 3$ meV, $\Delta k=0.01$ \AA$^{-1}$ (as reported in Ref.\ \cite{PHlawenkaNATCOMM2018}). 
Furthermore, we included band folding as described in Ref.\ \cite{PHlawenkaNATCOMM2018} for the simulation presented in the lower panel. 
Despite the low-velocity Dirac fermions we started with, both simulations give the appearance of high-velocity states at the Fermi level that  reproduce the ARPES experimental data presented in Refs.\ \cite{JJiangNATCOMM2013} and \cite{PHlawenkaNATCOMM2018}. 
(c) Two different ARPES intensity maps are reproduced from Refs.\ \cite{JJiangNATCOMM2013} and \cite{PHlawenkaNATCOMM2018} for direct comparison with our mixed-termination simulations in panel (b). (d) 
Adding electrons increases the Fermi level by a large amount due to the high velocity of the surface states above the chemical potential, whereas removing electrons decreases the Fermi level by only a small amount given the low surface state velocity below the chemical potential..
}
\label{fig:ARPESsim}
\end{figure*}

\subsection{Termination-dependent band bending}
In general, the surface termination can cause a redistribution of charge that affects the local electronic structure, an effect well studied in conventional semiconductors \cite{ZZhangChemRev2012}.  In bulk \smbp, Sm atoms donate equal amounts of charge to the B$_6$ octahedra above and below them. However, on the \oto surface the Sm atoms are under-coordinated; the B layer beneath the topmost Sm layer cannot accept all of the excess electrons, so they accumulate on the surface. This charge accumulation is qualitatively captured in our calculations of the electron transfer, which use Bader analysis to partition the DFT charge density (Fig.\ \ref{fig:charge-transfer}). 

The increased electron density near the \oto surface leads to reduced surface charge transfer shown as a blue line in Fig. \ref{fig:charge-transfer}(b), greater filling of the Sm orbitals, and to a slight downward bending of the surface bands.
On the other hand, Sm atoms at the \tto surface can donate a greater fraction of their electrons to the B layer below, because there are only half as many Sm atoms at the surface as in the bulk. Correspondingly, we found only a minor deviation in the calculated charge transfer at the \tto surface, shown as a red line in Fig.\ \ref{fig:charge-transfer}(b).
Although our Bader charge analysis quantitatively departs from the experimental Sm valence of around $+2.5$~\cite{tarasconJPhysFrance1980}, it provides a qualitative understanding of the charge transfer on the \smb surface.

To experimentally determine the accumulation of surface charge, we measured local differential conductance,  $dI/dV(\mathbf{r}, E)$, where $I$ is the tunneling current and $V$ is the bias applied to the sample with respect to the tip. On a typical ordered domain, there are three pronounced spectral features: a peak around $-150$ meV, a peak just below $E_F$, and a shoulder around $40$ meV, as shown in Fig.\ \ref{fig:ARPESsim}(a). 
The two filled-state peaks predominantly reflect contributions from the Sm $4f$ states, as determined by previous STM and ARPES measurements, and by dynamical mean-field theory calculations \cite{MYeeARXIV2013, pirieNatPhys2020, ZSunPRB2018, JDenlingerARXIV2013}. \black{Although the peak energies are homogeneous within each ordered domain [see Fig. \ref{fig:surface}(d-e)],} we found that the peaks are shifted downward on the \oto surface by about 20 meV compared to the \tto surface.

\subsection{Spectral function simulation}

ARPES spectra can be broadened by local band bending if the spot size encompasses multiple surface domains of different polarity. We investigated this possibility in \smb by deriving a spectral function for each termination, from our STM measurements \cite{pirieNatPhys2020}.  In accordance with our data, our simulation includes low-velocity Dirac states close to the chemical potential, connecting a light bulk $d$ band to two heavy bulk $f$ bands.
Each state includes a Fermi-liquid-like quasiparticle decay rate $\propto \omega^2$ \cite{CMVarmaPhysRep2002}. 
We simulated each termination by adjusting the energies of the $f$ and $d$ bands to match our STM spectra. Specifically, in the \oto spectral-function simulation, the occupied states are shifted down by 20 meV relative to the \tto simulation. We simulated ARPES spectra by computing an equal-weighted average of the spectral functions for each surface, then convolving the result with a Gaussian kernel that accounts for detector resolution and temperature broadening as shown in right panels of Fig.\ \ref{fig:ARPESsim}(b). Specifically, we mimic the detectors in Ref.\ \cite{JJiangNATCOMM2013} with parameters $T=12$ K, $\Delta E = 10 $ meV, and $\Delta k = 0.04 \mathrm{~\AA}^{-1}$, and Ref.\ \cite{PHlawenkaNATCOMM2018} with parameters $T = 1$ K, $\Delta E = 3$ meV, and $\Delta k = 0.01\mathrm{~\AA}^{-1}$.  In each case, our simulation captures the main features of the measured ARPES spectra as reproduced in Fig.\ \ref{fig:ARPESsim}(c): an apparent hybridization gap of approximately 20 meV, and in-gap surface states with an apparent high velocity, which seem to extrapolate to a buried Dirac point \footnote{The $\Gamma_8^{(2)}$ crystal-field-split $4f$ state is missing in our STM-derived simulation, but present in ARPES experiments. The discrepancy arises because STM does not couple strongly to the $\Gamma_8^{(2)}$ state, which lacks the correct symmetry to hybridize. However the lack of hybridization also means that the $\Gamma_8^{(2)}$ state does not play an important role in Kondo or topological physics.}.  

\section{Discussion}
\begin{figure}[t]
\centering
\includegraphics[width=0.95\columnwidth]{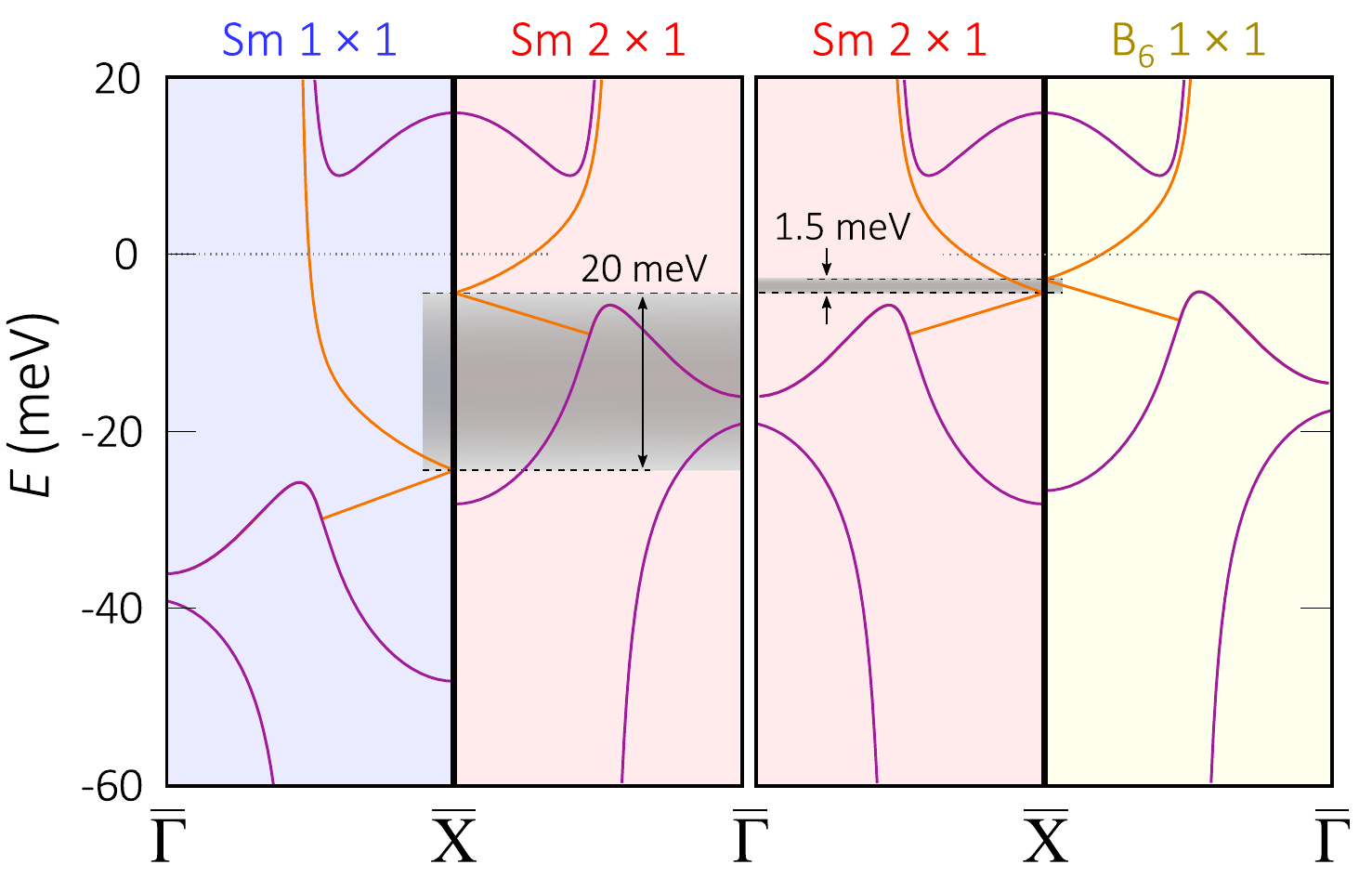}
\caption{\black{Band bending range on \smb surfaces. The Sm \oto is the most negatively charged surface with a measured downward band bending of 20 meV, compared to the charge neutral Sm \tto surface. Bands on the B$_6$ \oto surface, which is the most positively charged surface, are shifted up by 1.5 meV \cite{JiaoNatComm2016}. Therefore, our simulation including just the Sm \oto and Sm \tto surface spans more than 90 $\%$ of the maximum energy range of 21.5 meV.}} 
\label{fig:band-bending-range}
\end{figure}

A complete understanding of the cleaved \smb surface requires considering both electron-rich surfaces, such as the Sm \oto surface, and electron-deficient surfaces, such as the B-rich terminations. 
Importantly, our STM measurements have shown that the heavy Dirac surface states become flat only below the chemical potential \cite{pirieNatPhys2020}, leading to a highly electron-hole-asymmetric band-bending scenario, as depicted in Fig.\ \ref{fig:ARPESsim}(d) and Fig. \ref{fig:band-bending-range}. 
In such a scenario, we expect that surplus electrons, as found on Sm \oto terminations, primarily populate the steeper (upper) part of the surface-state dispersion [see Fig.\ \ref{fig:cartoon}(b)], producing a notable downward shift of spectral features, as shown in Fig.\ \ref{fig:ARPESsim}(a). 
Conversely, a surface deficient of electrons, as expected for B-rich terminations, would depopulate the very flat (lower) part of the surface-state dispersion. 
\black{Due to the dramatic difference in band slope (velocity) above and below the Fermi level, spectral features would be shifted upward by much less on a surface with missing electrons, than they would be shifted downward on a surface with the same number of excess electrons. Indeed, on B$_6$ \oto surfaces, STM measured a prominent peak at $-6.5$ meV \cite{JiaoNatComm2016}, which is shifted upward by only 1.5 meV compared to the corresponding peak on the neutral Sm \tto surface [see Fig. \ref{fig:ARPESsim}(a)]. Thus, the total band-bending range, defined by the most negatively charged Sm \oto termination and the most positively charged B$_6$ \oto termination, is 21.5 meV, as shown in Fig. \ref{fig:band-bending-range}. Therefore, our ARPES simulation, using data from the two surfaces we observe, covers more than 90\% of the maximum possible surface band-bending.}


While our study focuses on the (001) surface, recent ARPES experiments also reported high-velocity surface states on the (110) and (111) surfaces \cite{ohtsuboNatCommun2019, JDenlingerARXIV2016}. These reports are surprising because both surfaces are nominally non-polar and hence are expected to host low-velocity Dirac states.  In fact, magnetothermoelectric studies have already indicated the presence of heavy metallic states on the (110) surface \cite{YLuoPRB2015}, contrary to the ARPES measurement.  Under closer inspection by STM, the (110) surface appears to be inhomogeneous on small length scales \cite{SRosslerPM2016}. The intense atomic-scale disorder may alter the local electronic environment and cause local charging, analogous to termination-dependent band bending on the (100) surface \cite{ZSunPRB2018}. This local charging would be averaged in ARPES measurements, possibly resulting in enhanced surface-state velocities, similar to our simulations on the (100) surface (Fig.\ \ref{fig:ARPESsim}).

Band bending on \smb may also affect the perception of the hybridization gap and explain the apparent discrepancy between its size, as reported by ARPES and STM.
ARPES generally reports 15-20 meV for the part of the hybridization gap below $E_\mathrm{F}$, as shown in Fig.\ \ref{fig:ARPESsim}(c) \cite{JDenlingerJPSJC2014, NXuPRB2013, NXuPRB2014, MNeupaneNATCOMM2013, JiaoNatComm2016, PHlawenkaNATCOMM2018}, while the \textit{full} gap, as measured by STM, is only 8-15 meV ~\cite{pirieNatPhys2020, WRuanPRL2014,ZSunPRB2018,JiaoNatComm2016}. In Fig.\ \ref{fig:ARPESsim}(b), our ARPES simulation shows a large gap below $E_\mathrm{F}$, of about 25 meV, despite arising from a band structure with a gap of only 15 meV on the non-polar surface, as measured by STM. Specifically, averaging over different surface terminations blurs the top of the bulk valence band, which introduces an apparent increase of the hybridization gap on the occupied side. 
The full impact of excess charge on the surface Kondo environment and \textit{d}-\textit{f} hybridization remains an open theoretical question \cite{VAlexandrovPRL2015}.

\section{Conclusion}
 \smb is a promising platform for devices that exploit correlated topological phases, but its cubic and polar structure give rise to small, charged surface domains, on which band bending may locally distort the Dirac surface states.
 Using STM spectroscopy, we investigated two distinct surface terminations and measured a band shift of about 20 meV between them. These measurements guided a simulation of ARPES spectra, which captures the essential experimental features of ARPES, but remains consistent with STM conclusions~\cite{pirieNatPhys2020}.  Our results suggest that band bending is most pronounced on Sm-rich terminations, motivating the development of new surface treatments or epitaxial-growth techniques such as molecular beam epitaxy to achieve a more uniform termination.  Control over the termination would allow the important correlated surface states to be tuned closer to the Fermi level, without introducing disorder through chemical doping, which would be advantageous for future applications~\cite{yongApplPhysLett2014}.

\section*{Acknowledgements}
We thank Jonathan Denlinger, Emile Rienks and Yun Suk Eo for enlightening discussions and Johnpierre Paglione, Xiangfeng Wang, Zachary Fisk and Dae-Jeong Kim for providing the samples. Experiments were supported by National Science Foundation DMR-1410480 and partially as part of the Center for the Advancement of Topological Semimetals, an Energy Frontier Research Center funded by the U.S. Department of Energy (DOE), Office of Science, Basic Energy Sciences (BES) through the Ames Laboratory under its Contract No. DE-AC02-07CH11358 (STM measurements). HP and MHH were funded by the Gordon and Betty Moore Foundation’s EPiQS Initiative through Grant GBMF4536. CEM is supported by the Swiss National Science Foundation under fellowships P2EZP2\_175155 and P400P2\_183890. JJP and WP acknowledge Sahar Pakdel for her contribution at the initial stages of the theoretical work and financial support from Spanish MINECO through Grant FIS2016-80434-P, the Fundaci\'{o}n Ramón Areces, the Mar\'{i}a de Maeztu Program for Units of Excellence in R\&D (MDM-2014-0377), the Comunidad Autónoma de Madrid through the Nanomag COST-CM Program (S2018/NMT-4321), and the European Union Seventh Framework Programme under Grant agreement No.\ 604391 Graphene Flagship. WP was funded by the CNPq Fellowship programme (P\'{o}s-doutorado j\'{u}nior) under grant 405107/2017-0 and acknowledges the computer resources and assistance provided by the Centro de Computaci\'{o}n Cient\'{i}fica of the Universidad Aut\'{o}noma de Madrid and the computer resources at MareNostrum and the technical support provided by Barcelona Supercomputing Center (FI-2019-2-0007).

\end{document}